\documentstyle[preprint,aps]{revtex} 
\begin{document}
\draft
\preprint{\today}
\title{Phase transitions in BaTiO$_3$ from first principles}

\author{W.~Zhong and David Vanderbilt}

\address{Department of Physics and Astronomy,
  Rutgers University, Piscataway, NJ 08855-0849}

\author{K.~M.~Rabe}

\address{Department of Applied Physics, Yale University, New Haven,
CT 06520}
\date{\today}
\maketitle

\begin{abstract}
We develop a first-principles scheme to study ferroelectric phase
transitions for perovskite compounds.
We obtain
an effective Hamiltonian which is fully specified by first-principles
ultra-soft pseudopotential calculations.
This approach is applied to BaTiO$_3$, and
the resulting Hamiltonian is studied
using Monte Carlo simulations.
The calculated phase sequence, transition temperatures, latent heats,
and spontaneous polarizations are all in good agreement with
experiment.
The order-disorder vs.\ displacive character of the transitions
and the roles played by different interactions are discussed.
\vskip 0.2truein\noindent
{\sl Submitted to Phys.\ Rev.\ Lett.}
\vskip 0.5truein
\end{abstract}
\pacs{77.80.Bh, 61.50.Lt, 64.60.Cn, 64.70.-p}
\narrowtext

Because of their simple crystal structure, the cubic ferroelectric
perovskites present a special opportunity for the development of a
detailed theoretical understanding of the ferroelectric phase
transition.  However, even in BaTiO$_3$, a much-studied
prototypical example of this class of compounds \cite{lines}, many
aspects of the phase behavior are far from simple.
BaTiO$_3$ undergoes a succession of phase transitions, from the
high-temperature
high-symmetry cubic perovskite phase (Fig.\ \ref{stru}) to slightly
distorted ferroelectric structures with
tetragonal, orthorhombic and rhombohedral symmetry.
There is increasing evidence that the cubic-tetragonal
transition, at first thought to be of the simple displacive kind,
may instead be better described as of the order-disorder type.

A comparison with the related cubic perovskites indicates that
this and other aspects of the phase transformation behavior in
BaTiO$_3$ are not universal, but rather must depend on details
of the chemistry and structural energetics of the particular
compound.
Therefore, it is of the first importance to develop
a microscopic theory of the materials properties which determine
the ordering of the phases, the character and thermodynamic order
of the transitions, and the transition temperatures.
The value of a microscopic approach has long been appreciated, but
its realization was hindered by the difficulty of determining
microscopic parameters for individual compounds.
The forms of phenomenological model Hamiltonians
\cite{lines,dove,pytt,cowl}
were limited by
the available experimental data, leading to oversimplification and
ambiguities in interpretation.
For the perovskite oxides, empirical \cite{bilz} and nonempirical
pair potential methods \cite{boyer} did not offer the
high accuracy needed for the construction of realistic models.
Recently, high quality first-principles calculations within the
local density approximation (LDA)
have been shown to provide accurate total-energy surfaces for
perovskites\cite{cohen,singh,king1,king11}.
While an {\it ab-initio} molecular-dynamics simulation of the
structural
phase transition is not computationally feasible at present,
the application of these first-principles methods can clearly form
a foundation for the realistic study of the finite-temperature
phase transitions.

In this paper, we pursue a completely first-principles approach
to study the ferroelectric phase transitions
in BaTiO$_3$.  In particular, we (i) construct an effective
Hamiltonian to describe the important degrees of freedom of the
system
\cite{rabe,rabeuw}, (ii) determine all the parameters of this
effective
Hamiltonian from high-accuracy {\it ab-initio} LDA calculations
\cite{king1,vand1,zhong}, and
(iii) carry out Monte Carlo (MC) simulations to determine the phase
transformation behavior of the resulting system.
We find the
correct succession of phases, with transition
temperatures and spontaneous polarization in reasonable agreement
with experiment. Strain coupling is found to be crucial in producing the
correct succession of low-symmetry phases.
Finally, by analyzing
the local distortions and phonon softening, we find
the cubic-tetragonal transition in BaTiO$_3$
to be intermediate between the displacive and order-disorder limits.

Briefly, the effective Hamiltonian is constructed as follows.
Since the ferroelectric transition involves only small
structural distortions,
we represent the energy surface by a Taylor expansion
around the high-symmetry cubic perovskite structure,
including fourth-order anharmonic terms. Because the contribution
to the partition function decays exponentially with increasing
energy,
we simplify this expansion by including only low energy distortions.
Among all the possible phonon excitations,
the long-wavelength acoustic modes (strain) and lowest
transverse-optical phonon modes (soft
modes) have the lowest energy.  It is therefore our
approximation to include only these two kinds of phonon excitations,
thus reducing the number of degrees of freedom
per unit cell from fifteen to six.
This approximation could later be systematically improved, or
entirely
removed, by including higher-energy phonons.

It is straightforward to describe the strain degrees of freedom
associated with the acoustic modes in terms of displacement vectors
${\bf v}_l$ associated with each unit cell $l$.  In a similar manner,
we
introduce variables ${\bf u}_l$ to describe the amplitude of a
``local
mode'' associated with cell $l$.  The properly chosen local mode
should reproduce the soft-mode phonon dispersion relation
throughout the Brillouin zone, preserve the
symmetry of the crystal, and minimize interactions between adjacent
local
modes. The local mode chosen for BaTiO$_3$ is shown in Fig.\
\ref{stru}.
The terms in our Taylor expansion of the energy in the variables
$\{ {\bf u}  \}$ and $\{ {\bf v}  \}$ are organized as follows:
{\it (i)}
a soft-mode self-energy $E^{\rm self}(\{ {\bf u} \})$ containing intrasite
interactions to quartic anharmonic order;
{\it (ii)} a long-range dipole-dipole coupling $E^{\rm dpl}(\{ {\bf
u}  \})$ and a short-range (up to third neighbor) correction $E^{\rm
short}(\{ {\bf u}  \})$ to the intersite coupling, both at harmonic
order;
{\it (iii)} a harmonic elastic energy $E^{\rm elas}(\{ {\bf v}  \})$;
and {\it (iv)} an anharmonic strain--soft-mode coupling
$E^{\rm int}(\{ {\bf u}  \},\{ {\bf v}  \})$ containing
Gruneisen-type interactions (i.e., linear in strain and quadratic
in soft-mode variables).
The cubic symmetry greatly reduces the number of expansion
coefficients
needed.  All the expansion parameters are determined from
highly-accurate
first-principles LDA calculations
applied to
supercells containing up to four primitive cells (20 atoms).
The calculation of the needed microscopic parameters within LDA
for BaTiO$_3$ has been made possible by the use
of Vanderbilt ultra-soft pseudopotentials \cite{vand1},
which make large-scale calculations tractable
at the high level of accuracy needed,
and by the recent theory of polarization
of King-Smith and Vanderbilt\cite{king2}, which provides a
convenient method of calculating the dipolar interaction
strengths \cite{zhong}.
The details of the Hamiltonian, the first-principles calculations,
and
the values of the expansion parameters will be reported elsewhere
\cite{zhong2}.

We solve the Hamiltonian using Metropolis Monte Carlo
simulations\cite{metro,MC}
on an $L\times L \times L$ cubic
lattice with
periodic boundary conditions.  The homogeneous part of the strain in
the system is separated out and treated as six extra degrees of
freedom.
Since most energy contributions (except $E^{\rm dpl}$) are local, we
choose the
single-flip algorithm and define one Monte Carlo sweep (MCS) as $L^3$
flip attempts.

The ferroelectric phase transition is very sensitive to hydrostatic
pressure, or equivalently, to lattice constant.  The LDA-calculated
lattice constants are typically 1\% too small, and
even this
small error can lead to large
errors in the zero-pressure
transition temperatures.
The effect of this systematic error can largely be
compensated by exerting a negative pressure that expands the lattice
constant to the experimental value.  For BaTiO$_3$, we choose
$P=-4.8$ GPa which gives the best overall agreement for the
computed volumes for the four phases with their experimental values.
The following
simulations and analysis are for this pressure.

In our simulation, we concentrate on identifying the succession of
different phases, determining the phase transition
temperatures, and extracting qualitative
features of the transitions. We also focus on identifying the
features of the Hamiltonian which most
strongly affect
the transition properties.
For these purposes, it is most convenient to monitor directly the
behavior of the order parameter.  In the case of the ferroelectric
phase transition, this is just the polarization vector (or
equivalently,
the soft-mode amplitude vector $\bf u$) averaged over the simulation
cell.
To avoid effects of possible rotation of the polarization vector and
to identify the different phases clearly,
we choose to accumulate the absolute values of the largest, middle,
and smallest components of the
averaged local-mode vector for each step, denoted by $u_1$, $u_2$,
and $u_3$,
respectively ($u_1 > u_2 > u_3$).  The
cubic (C), tetragonal (T), orthorhombic (O), and rhombohedral (R)
phases
are then characterized by zero, one, two, and three non-zero
order-parameter
components, respectively.
As a reference, the average local mode
amplitude $u = \sum_i |{\bf u}_i| / N $ is also monitored.
Here, ${\bf u}_i$ is the
local mode vector at site $i$ and $N$ is the total number of
sites.

Fig.~\ref{u-T} shows the quantities $u_1$, $u_2$, $u_3$, and $u$
as functions of temperature
in a typical simulation for an $L=12$ lattice.
For clarity, we show only the cooling down process.
The values are averaged over 7000 MCS's after the system
reaches equilibrium, so that the typical fluctuation of order
parameter components is less than 10\%.
We find that
$u_1$, $u_2$, and $u_3$ are all very close to zero at high
temperature.  As the
system cools down past 340K, $u_1$ increases and becomes
significantly larger than $u_2$ or $u_3$.  This indicates the
transition to the tetragonal phase.  The homogeneous-strain
variables confirm that the shape of the simulation cell becomes
tetragonal.  Two other phase transitions occur as the temperature
is reduced further.  The T--O transition occurs at 255K (sudden
increase of $u_2$) and the O--R transition occurs at 210K (sudden
increase of $u_3$).  The shape of the simulation cell also shows
the expected changes.  The sequence of transitions exhibited by the
simulation is the same as that observed experimentally.

The transition temperatures are located by careful cooling and
heating sequences.  We start our simulation at a high
temperature and equilibrate in the cubic phase. The temperature is
then reduced in small steps.  At each temperature, the system is
allowed to relax for 10,000 MCS (increased to 25,000 and then to
40,000 MCS's close to the transition). After each transition is
complete, the system is reheated slowly to detect any possible
hysteresis.  The calculated transition temperatures are shown in
Table \ref{table1}.  Simulations for three lattice sizes are
performed; the error estimates in the table reflect the hysteretic
difference between cooling and heating, which persists even after
significant increase of the simulation time.  (The C--T transition
temperature for $L=10$ is difficult to identify because of large
fluctuations between phases.) The calculated transition temperatures
are well converged with respect to system size, and are in good
agreement with experiment.
The saturated spontaneous polarization $P_{\rm s}$ in different phases can be
calculated from the average local mode variable. The results are also shown
in Table \ref{table1}.  We find almost no finite-size effect,
and the agreement with experiment is very good for O and T phases.
The disagreement for the R phase may be due in part to twinning
effects in the experimental sample \cite{wied}.

One way to determine the order of the transition is to calculate the
latent heat. An accurate determination of the latent heat would
require
considerable effort; here, we only try to provide good estimates.
We approach the transition from both high-temperature
and low-temperature sides until the point is reached where both
phases
appear equally stable.  The difference of the average total energy
is then the latent heat\cite{latent}. This estimate should be good as long as
some hysteresis is
present.  The calculated latent heat (Table \ref{table1}) is in rough
agreement with the rather scattered experiment data.
The discrepancy for the C--T transition can be partly attributed to
the finite-size effect.
We find that, taking into account finite-size effects, the latent heats
for all three transitions are significantly non-zero,
suggesting all transitions are first-order.
For the T--O and O--R transitions, this is consistent with
Landau theory, which requires a transition to be first-order
when the subgroup relation does not hold between the symmetry groups
below and above $T_c$.
For the C--T transition, although it has the largest latent heat,
we find indications that it is
the most weakly first order.  The relatively large finite-size effect
suggests a relatively long correlation length near the transition,
and the change of the order parameter during the transition is
also more gradual (Fig.\ref{u-T}).

Next, we investigate the extent to which the cubic-tetragonal
transition can be characterized as order-disorder or displacive.
In real space, these possibilities can be distinguished by inspecting
the distribution of the local-mode vector ${\bf u}_i$ in the cubic
phase just above the transition.
A displacive (microscopically nonpolar) or order-disorder
(microscopically
polar) transition should be characterized by a single-peak or
double-peak structure, respectively.
The distribution of $u_x$ at $T=400K$ is shown in Fig.~\ref{udos}.
It exhibits a rather weak
tendency to a double-peaked structure, indicating a transition which
has some degree of order-disorder character.
We also see indications of this in the $u-T$ relation
in Fig.\ref{u-T}; even in the cubic phase, the magnitude of the
local-mode vector
$u$ is significantly non-zero and close to that of the rhombohedral
phase.
Although the components of the local modes change dramatically
during the phase transition, $u$ only changes slightly.

In reciprocal space, a system close to a displacive
transition should show large and strongly temperature-dependent
fluctuations of certain phonons (soft modes) confined to a small
portion of the Brillouin zone (BZ).  For an extreme
order-disorder transition, on the other hand, one expects the
fluctuations to be distributed over the whole BZ.
For BaTiO$_3$, we calculated the average Fourier modulus of the
soft TO mode $<|u(q)|^2>$ at several temperatures
just above the C--T transition.
A strong increase of $<|u(q)|^2>$ as $T\rightarrow T_c$
would indicate phonon softening.
As expected, we do observe this behavior for modes at
$\Gamma$.  While these modes become
``hard'' rather quickly along most directions away from $\Gamma$,
they remain soft at least half-way to the BZ boundaries along the
$\{100\}$ directions, again indicating some order-disorder character.
Thus, from the
example of BaTiO$_3$, it seems that a positive on-site quadratic
coefficient
does not automatically imply a displacive character for the
transition.
Rather, the relevant criterion is the extent to which the
unstable phonons extend throughout the BZ.

Our theoretical approach allows us to investigate the roles
played by different types of interaction in the phase
transition.  First, we study the effect of strain.  Recall
that the strain degrees of freedom were separated into local and
homogeneous parts, representing finite- and
infinite-wavelength acoustic modes, respectively.  Both parts
were included in the simulations.  If we eliminate the
local strain (while still allowing homogeneous strain), we
find almost no change in the transition temperatures.  This
indicates that the effect of the short-wavelength acoustic
modes may not be important for the ferroelectric phase
transition.  If the homogeneous strain is frozen, however, we
find a direct cubic--rhombohedral phase transition, instead
of the correct series of three transitions.  This
demonstrates the important role of homogeneous strain.
Second, we studied the significance of the long-range Coulomb
interaction in the simulation. To do this, we changed the
effective charge of the local mode (and thus the
dipole-dipole interaction), while modifying other parameters
so that the frequencies of the zone-center and zone-boundary
phonons remain in agreement with the LDA values.  We found
only a slight change (10\%) of the transition temperatures
when the dipole-dipole interaction strength was reduced by
half, but a further reduction changed the results
dramatically (in fact the ground state becomes
a complex antiferroelectric structure). This result
shows that it is essential to include the long-range
interaction, although small inaccuracies in the calculated
values of the effective charges or dielectric constants may
not be very critical.  On the other hand, our tests do
indicate a strong sensitivity of the $T_c$'s to any deviation
of the fitted zone-center or zone-boundary phonon frequencies
away from the LDA results.  Thus, highly accurate LDA
calculations do appear to be a prerequisite for an accurate
determination of the transition temperatures.

Our approach opens several avenues for future study.
Allowing a higher-order expansion of the  energy surface
might allow an accurate determination of the phase diagram.
More extensive
Monte-Carlo simulations on larger systems, and with careful
analysis of finite-size scaling, could provide more precise
transition temperatures, free energies, and latent heats\cite{lee}.
Finally, the theory would be more satisfying if the 1\%
underestimate of the lattice constant in the LDA calculation
could be reduced or eliminated.

In conclusion, we have developed a first-principles approach
to the study of structural phase transitions and the
calculation of transition temperatures in BaTiO$_3$.  We have
obtained the transition sequence, transition temperatures, and
spontaneous polarizations, and found them to be in good agreement
with experiment. We find that long-wavelength acoustic modes and
long-range dipolar interactions both play an important role
in the phase transition, while short-wavelength acoustic
modes are not as relevant.  The C--T phase transition is not
found to be well described as a simple displacive transition;
on the contrary, if anything it has more order-disorder character.

We would like to thank R.D. King-Smith, U. V. Waghmare, R. Resta, Z.~Cai,
and A.M.~Ferrenberg for useful discussions. This
work was supported by the Office of Naval Research under contract
numbers N00014-91-J-1184 and N00014-91-J-1247.

\newpage
\begin{table}
\caption{Calculated transition temperatures $T_{\rm c}$,
saturated spontaneous polarization P$_{\rm s}$, and estimated
latent heat $l$, as a function of simulation cell size.
\label{table1}}
\begin{tabular}{l|c|cccc}
                &phase & $L=10$   & $L=12$ & $L=14$ & expt\tablenotemark[1]\\
\hline
                & O--R & 210$\pm$10 & 220$\pm$10 & 218$\pm$2 & 183 \\
$T_{\rm c}$ (K) & T--O & 252$\pm$2  & 260$\pm$5  & 264$\pm$1 & 278 \\
                & C--T & $\sim$320     & 341$\pm$1  & 342$\pm$2 & 403 \\
\hline
                & R    & 0.43       & 0.43       &  0.43     & 0.33 \\
P$_{\rm s}$ (C/m$^2$) & O & 0.36    & 0.35       &  0.36     & 0.36 \\
                & T   &     0.28    & 0.28       &  0.28     & 0.27 \\
\hline
%
                & O--R &  58        &    50      &  50     & 33--60 \\
$l$ (J/mol)     & T--O & $\ge$86    &    92      & 100       & 65--92 \\
                & C--T & --         & $\ge$73    & 115       & 196--209 \\
\end{tabular}
\tablenotetext[1]{T. Mitsui {\em et al.}, {\sl  Landolt-Bornstein
numerical data and functional relationships in science and
technology} (Springer-Verlag, 1981), NS, III/{\bf 16}.}

\end{table}

\begin{figure}
\caption{The structure of cubic perovskite compounds BaTiO$_3$. Atoms
Ba, Ti and O are represented by shaded, solid, and empty circles
respectively. The areas of the vectors indicate the magnitudes of
the displacements for a local mode polarized along $\hat{\bf x}$.
 \label{stru} }
\end{figure}

\begin{figure}
\caption{The averaged largest, middle, and smallest components
$u_1$, $u_2$, $u_3$ and amplitude $u$ of local modes as a function of
temperature in a
cooling-down simulation of a $12\times 12\times 12$ lattice.
The dotted lines are guides to the eyes. \label{u-T}}
\end{figure}

\begin{figure}
\caption{The distribution of a Cartesian component of the local mode
variable in the cubic phase at $T=400K$. \label{udos}}
\end{figure}


\begin{references}
\bibitem{latent}
Under a {\it physical} applied pressure, the latent heat would
be the change of enthalpy, not energy.  If this were applied to
our case of a {\it fictitious} pressure, the calculated latent
heats would be approximately twice as large.
\end{references}
\end{document}